\documentclass{webofc}
\usepackage[varg]{txfonts}   % Web of Conferences font
\usepackage{lineno}     % for line numbering
\usepackage[linktocpage=true]{hyperref}
\hypersetup{
    colorlinks=true,
    linkcolor=cyan,
    filecolor=cyan,  
    citecolor=cyan, 
    linkcolor=cyan,    
    urlcolor=cyan,
    pdftitle={Overleaf Example},
    pdfpagemode=FullScreen,
    }
\begin{document}
\modulolinenumbers[2]
% \pagewiselinenumbers
\switchlinenumbers   % allow to put line numbers on the outer margins
%\linenumbers
%
\title{A multi-differential investigation of strangeness production in pp collisions with ALICE}

\author{\firstname{Romain} \lastname{Schotter}\inst{1,2,3}\fnsep\thanks{\email{romain.schotter@cern.ch}}, for the ALICE collaboration.
        % etc.
}

\institute{Institut Pluridisciplinaire Hubert Curien, UMR7178, CNRS -- Strasbourg, France
\and Université de Strasbourg -- Strasbourg, France
\and EUR QMat -- Strasbourg, France}

\abstract{%
  In these proceedings, two multi-differential analyses performed in pp collisions collected by the ALICE collaboration during the LHC Run 2 are presented. One investigates the dependence of strange particle production with multiplicity and \textit{effective} energy, whereas the other clarifies how strangeness enhancement is correlated to the leading jet in the event. The results suggest that strangeness production at the LHC depends strongly on effective energy, and originates dominantly from the transverse region with respect to the leading jet direction.
}
\maketitle
\section{Introduction}
\label{intro}
By colliding heavy nuclei at extremely high energy, a new state of nuclear matter, in which quarks and gluons are deconfined --- the quark--gluon plasma (QGP) ---, is formed. One of its historical key signatures is the \textit{strangeness enhancement} \cite{RefRafelski}, which consists in the increase of the relative yields of multistrange particles, such as $\Xi\rm(dss)$ and $\Omega\rm(sss)$, to non-strange hadrons. Recently, it has been observed that such yield enhancement also scales smoothly with the charged particle multiplicity in proton--proton (pp) collisions \cite{RefStrangenessEnhancement}. The presence of this phenomenon in such a small system questions the very foundations of the QGP concept. 

Several phenomenological models, based on different approaches and mechanisms, are being developped but none of them has been able to provide an unambiguous explanation so far \cite{RefMCComparison}. Further experimental inputs are required in order to distinguish them, and this is achieved via more multi-differential studies related to the strange hadron production.

One can perform an analysis in order to separate the contribution of initial-state effects on the strangeness enhancement from the final ones. Indeed, the distribution of the charged particle multiplicity is a characteristic of the final state of the collision, but it also depends on the energy effectively available for particle production, which is related to the initial interactions. The idea is to investigate the dependence of $\Xi$ baryon production multi-differentially in multiplicity and effective energy. Similarly, the relative contribution of hard processes --- such as jets --- and softer processes --- occuring in transverse region to the jet axis --- to the production of $\Xi$ baryons can be now studied more differentially.

%here the strange particles of interest are the $\rm K_{S}^{0}$ and $\Xi$.

\section{Detector setup and data sample}
\label{sec:Detector}
%; for $\rm K_{S}^{0}$, their V0 decay channel is used : $\rm K_{S}^{0} \rightarrow \pi^{+} \pi^{-} $ (69.2\%)

In these two analyses, the $\Xi$ baryons are studied in their cascade decay channel: $\Xi^{\pm} \rightarrow \pi^{\pm} \Lambda \rightarrow \pi ^{\pm} \pi ^{\pm} p^{\mp}$ (63.9\%). They are reconstructed at midrapidity ($|y| < 0.5$), using the central-barrel detectors of the ALICE experiment \cite{RefALICEExp}: the Inner Tracking System (ITS) and the Time Projection Chamber (TPC) for the track reconstruction and particle identification. At forward pseudorapidity ($\eta$), charged particle multiplicity is determined with the V0M estimator, based on the sum of the signal amplitudes measured in the scintillator arrays, V0A ($2.8 < \eta < 5.1$) and V0C ($-3.7 < \eta < -1.7$). The number of clusters found in the two innermost layers of the ITS (SPD) can also provide a multiplicity estimation but at midrapidity ($|\eta| < 0.8$): this is the SPDCluster estimator. Finally, the \textit{ZDC Energy Sum} estimates the energy of particles produced at very forward rapidity ($6.5 < |\eta| < 7.4$ for protons, and $|\eta| > 8.8$ for neutrons) that have been deposited in the two Zero Degree Calorimeters (ZDC).

For the first analysis, the data sample is composed of $129 \times 10^6$ minimum bias events at a center-of-mass energy $\sqrt{s}$~=~13~TeV, recorded in 2015, 2017 and 2018. Concerning the second study, the dataset contains approximatively $420 \times 10^6$ events, coming from pp collisions at $\sqrt{s}$~=~13~TeV collected in 2016, 2017, and 2018. Additionally, $920 \times 10^6$ pp collisions at $\sqrt{s}$~=~5~TeV are used as well for comparison.
%For the first analysis, the data sampled is composed of 129M minimum bias events at $\sqrt{s}$=13~TeV, recorded in 2015, 2017 and 2018. Concerning the second study, the total dataset contains approximatively 370M events for the $\rm K_{S}^{0}$ analysis, and 420M events for the $\Xi$ analysis, both coming from pp collisions at $\sqrt{s}$=13~TeV collected in 2016, 2017, and 2018. A data sample of 920M pp collisions at $\sqrt{s}$=5~TeV is also used for comparison purposes.

\section{Dependence of strange particle production with multiplicity and effective energy}
\label{sec-3}

\subsection{Details on the effective energy and the double differential analysis}
\label{subsec:EffectiveEnergy}

In these proceedings, the effective energy is defined as the energy effectively available for particle production in the initial stages of the pp collision, and could be equal in principle to the center-of-mass energy. One has also to consider the baryons emitted in the very forward direction --- the so-called forward leading baryon emission~\cite{RefEffEnergy2007} ---, which carry a large fraction of the incident beam energy and therefore reduce the energy available for particle production at midrapidity. The energy carried by these baryons can be measured by the two ZDCs and thus let us define a proxy for the effective energy as $\sqrt{s}-\langle\rm ZDC \ energy \ sum\rangle$.

As mentionned in Section~\ref{intro}, multiplicity at midrapidity and effective energy are correlated~\cite{RefZDCVsMult}. In order to disentangle contributions from the initial (effective energy) and final (multiplicity) stages of the collision, the analysis is performed differentially in event multiplicity classes based on two estimators: SPDCluster and V0M. It has been observed that:
\begin{itemize}
\item[$-$] Setting the SPDCluster class sharply fixes the multiplicity at midrapidity, and for these classes, further selections on the V0M class allow to vary the effective energy.
\item[$-$] Conversely, fixing the V0M class constrains the effective energy to a narrower range, allowing further  SPDCluster selections to classify events with different multiplicity.
\end{itemize}

\subsection{Experimental results}
\label{subsec:ExpRes}

\begin{figure}[h]
\centering
\includegraphics[width=10cm,clip]{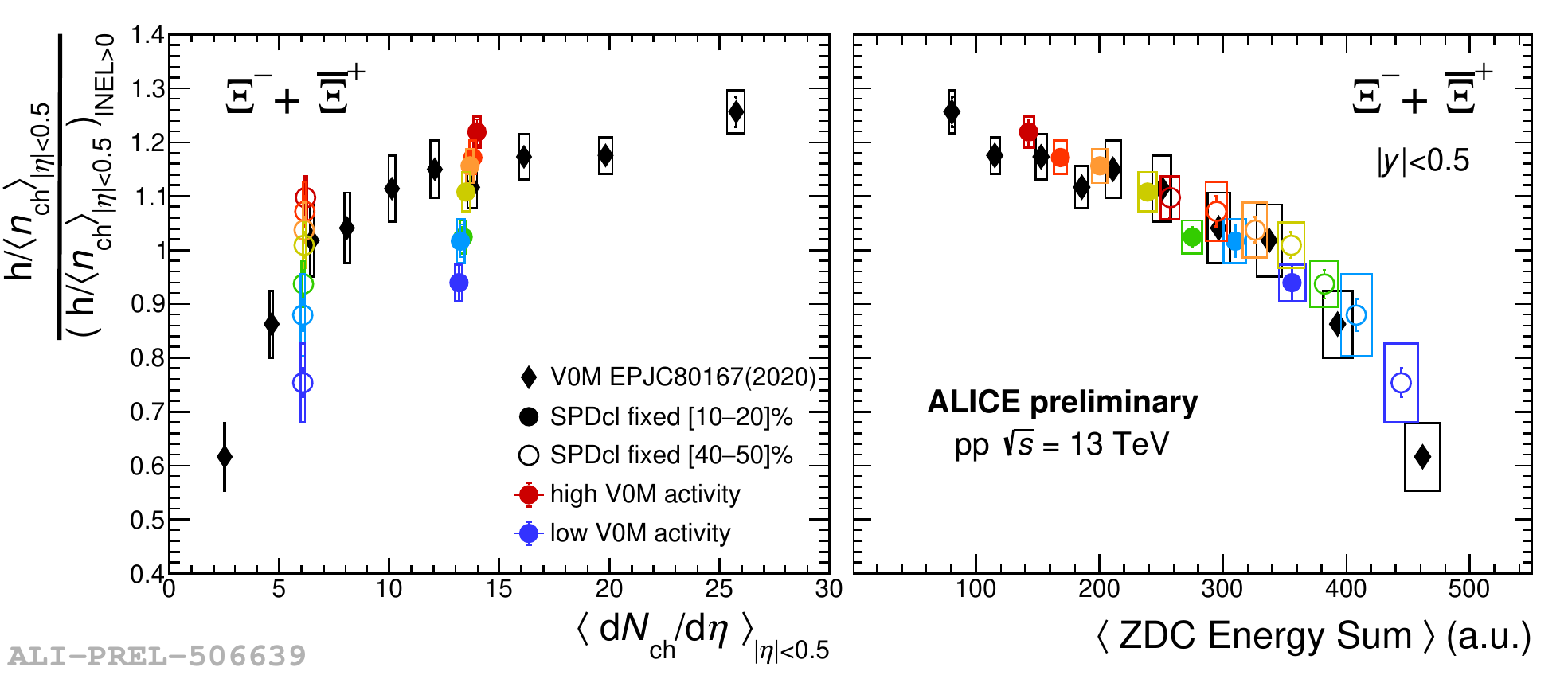}
\caption{$\Xi$ yield per charged particle self-normalized to INEL>0 as a function of multiplicity at midrapidity on the left, and of the ZDC energy sum on the right. The black diamonds correspond to the analysis performed in V0M classes alone \cite{RefOrdinarV0MAnalysis}; the colored full (open) squares present the double differential analysis with SPDCluster fixed in $[10,20]\%$ ($[40, 50]\%$).}
\label{fig:FixedMult}       % Give a unique label
\vspace*{-0.7cm}
\end{figure}

Figure~\ref{fig:FixedMult} shows the $\Xi$ yield per charged particle when multiplicity at midrapidity is fixed in the double differential analysis. The black diamonds correspond to different V0M multiplicity percentile class.  On the left panel, one can observe that the $\Xi$ yield increases with the charged particle multiplicity at midrapidity: this is the strangeness enhancement. On the right panel, the correlation between multiplicity and effective energy is clearly visible from the decrease (or increase) of the $\Xi$ yield with the average ZDC energy sum (or the effective energy).

The double differential analysis indicates that the $\Xi$ yield per charged particle increases at fixed multiplicity in the left panel of Fig.~\ref{fig:FixedMult}. From the right panel, one can see that this yield actually increases with effective energy. Moreover, the V0M standalone and the double differential measurement points are compatible within uncertainties, suggesting that the so-known strangeness enhancement with multiplicity is likely a dependence on effective energy.
 
\begin{figure}[h]
\centering
\includegraphics[width=10cm, clip]{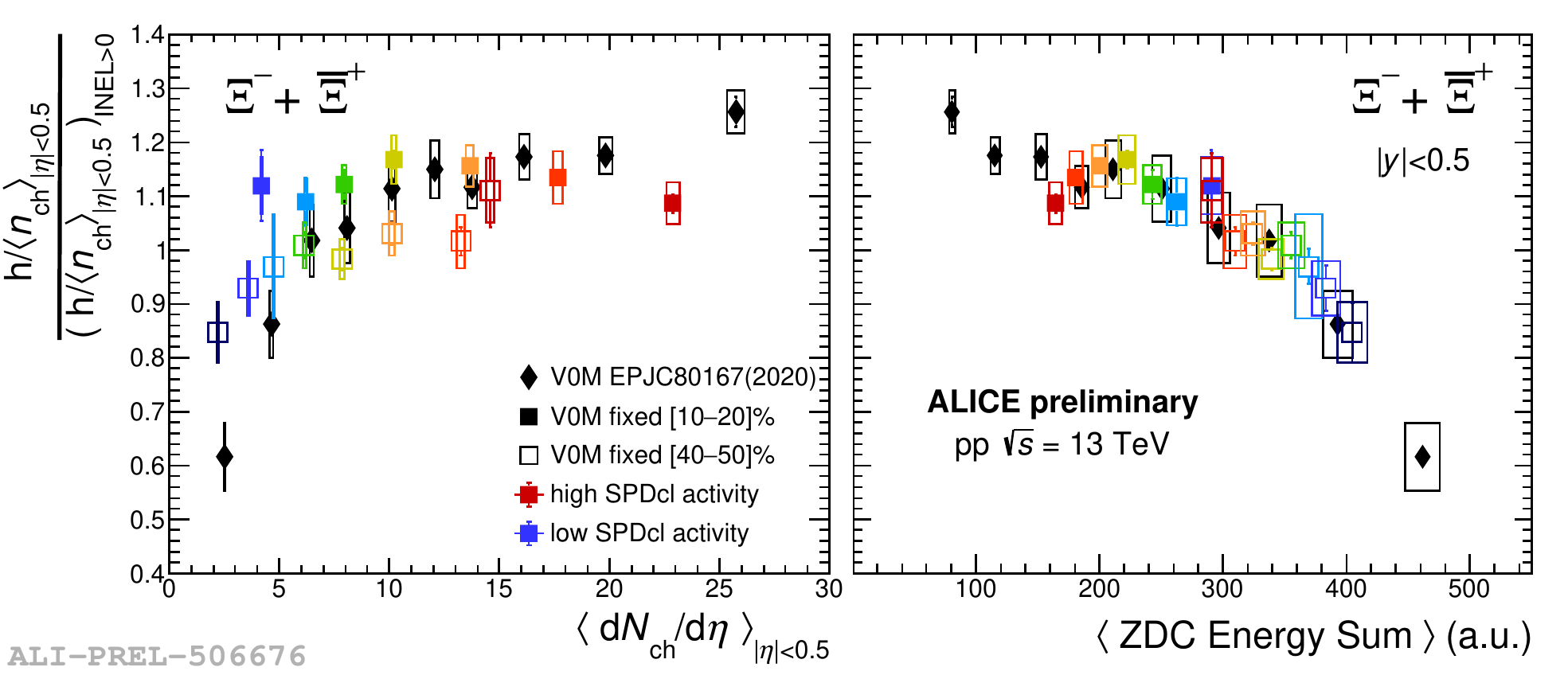}
\caption{$\Xi$ yield per charged particle self-normalized to INEL>0 as a function of multiplicity at midrapidity on the left, and of the ZDC energy sum on the right. The black diamonds correspond to the analysis performed in V0M classes alone \cite{RefOrdinarV0MAnalysis}; the colored full (open) circles present the double differential analysis with V0M fixed in $[10,20]\%$ ($[40, 50]\%$).}
\label{fig:ReducedEffEnergy}       % Give a unique label
\vspace*{-0.7cm}
\end{figure}

The situation with constrained effective energy is depicted in Fig.~\ref{fig:ReducedEffEnergy}. The right panel shows that effective energy has been divided into two ranges: one with a large and another one with a small average effective energy, represented in full and open markers, respectively. As can be seen on the left panel, the $\Xi$ yield shows an almost flat dependence with the charged particle multiplicity at midrapidity for the case of large effective energies. On the contrary, for smaller effective energies the yield increases rapidly at first and then the trend with multiplicity becomes milder. It seems that, with constrained effective energy, the rise of strangeness enhancement with multiplicity is strongly affected.

\section{Study of strange hadron production in the regions toward and transverse to the leading jet}
\label{sec:Jet}

\subsection{Angular correlation method}
\label{subsec:Method}

To separate strange hadrons produced towards the leading jet from the ones produced transverse to the leading jet, the angular correlation method is applied. It consists in forming pairs between the leading particle in the event, i.e. the primary charged particle with the highest $p_{\rm T}$ and $p_{\rm T} > 3$~GeV/$c$, and the particles of interest, e.g. the $\Xi$ baryons. The angular correlation is provided by the distributions of pairs in relative pseudorapidity ($\Delta \eta$) and azimuthal angle ($\Delta \varphi$) ; strange hadrons produced towards the leading jet originate from small ($\Delta \eta$, $\Delta \varphi$).

\subsection{Experimental results}
\label{subsec:ExpRes2}

%For two-column wide figures use syntax of figure~\ref{fig-2}
%\begin{figure*}
%\centering
%% Use the relevant command for your figure-insertion program
%% to insert the figure file. See example above.
%% If not, use
%\vspace*{5cm}       % Give the correct figure height in cm
%\label{fig-2}       % Give a unique label
%\end{figure*}

\begin{figure}[!htbp]
\vspace*{-0.4cm}
\centering
\sidecaption
\includegraphics[width=9cm,clip]{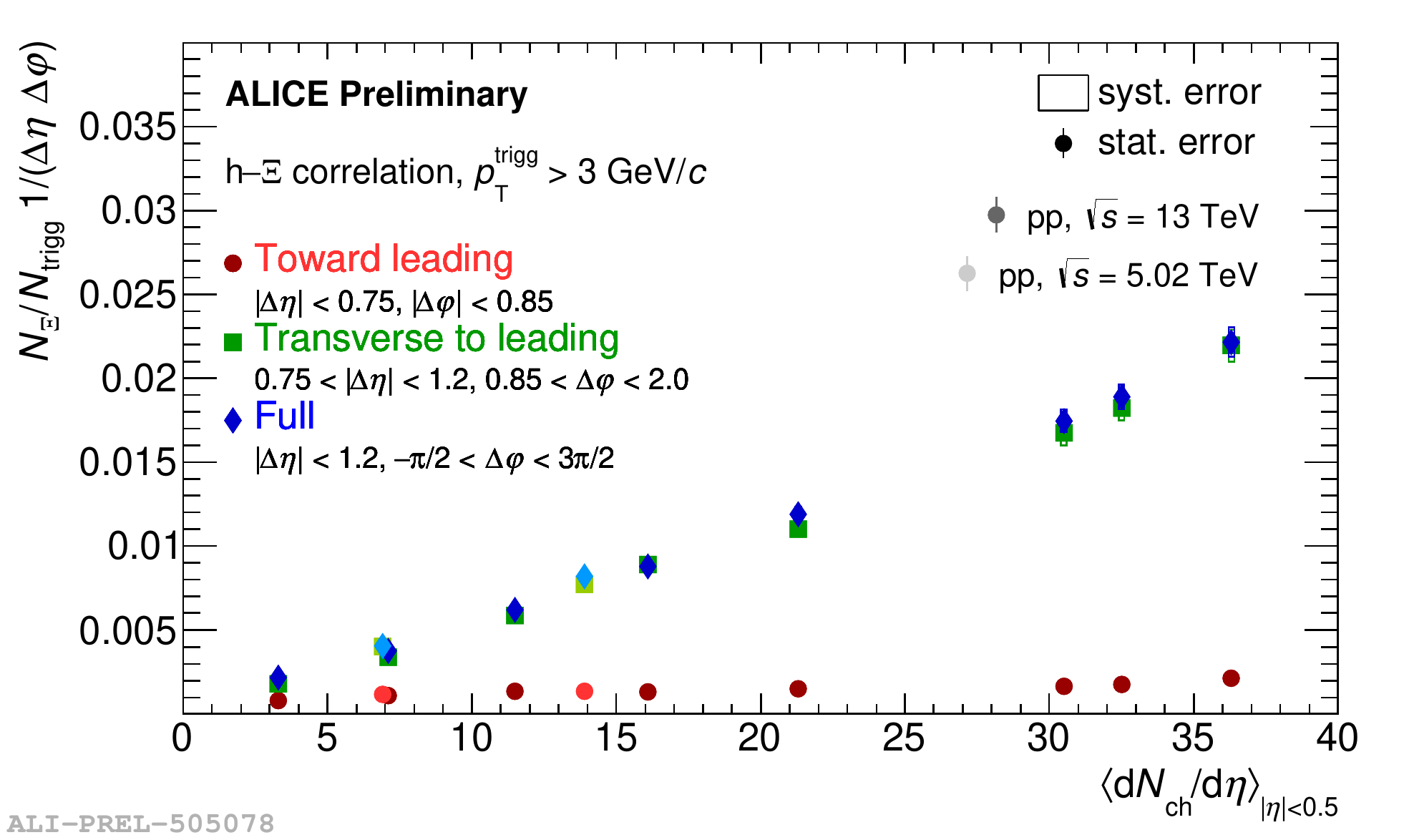}
\caption{Distribution of yield per trigger particle and per unit of ($\Delta \eta, \ \Delta \varphi$) of $\Xi$~baryons as a function of the charged particle multiplicity at midrapidity: full yield in blue diamonds, toward the leading jet yield in red circles, and transverse to the leading jet yield in green squares. Results for pp at $\sqrt{s}=13$ TeV are in dark colors, and pp at $\sqrt{s}=5$~TeV in light colors.}
\label{fig:XiYield}       % Give a unique label
\end{figure}

Figure~\ref{fig:XiYield} shows the $\Xi$ yield density per triggered event --- ratios are normalized to the considered ($\Delta \eta$, $\Delta \varphi$) region --- extracted in three ($\Delta \eta$, $\Delta \varphi$) regions as a function of the charged particle multiplicity at midrapidity in pp collisions. One can observe that both full and transverse to leading jet yields increase with the charged particle multiplicity at midrapidity, whereas the production in the toward-leading-jet region exhibits a quasi-flat dependence on multiplicity. This means that $\Xi$ are dominantly produced in the transverse region to the leading jet, and therefore the transverse to leading jet processes are the dominant contribution to the strangeness enhancement in pp collisions. It should be also noted that the measured yields in pp at $\sqrt{s}=13$ TeV are consistent with the ones at $\sqrt{s}=5$ TeV.

%A comparison of the strangeness enhancement between $\Xi$ and $\rm K_{S}^{0}$ in the toward and transverse to leading jet regions is plotted in figure~\ref{fig:YieldRatio}. It should be noted that the transverse to leading jet yield ratio of $\Xi$ to $\rm K_{S}^{0}$ increases with the multiplicity at mid-rapidity, and is compatible with the full yield ratio. In the toward leading jet region, the yield ratio is systematically smaller whatever the multiplicity value, but from the bottom panel, its trend with multiplicity is consistent with the one in the transverse to leading jet region. This means that $\Xi$ to $\rm K_{S}^{0}$ yield ratio in pp collisions is dominated by transverse to leading jet processes
%
%\begin{figure}[h]
%% Use the relevant command for your figure-insertion program
%% to insert the figure file.
%\centering
%\sidecaption
%\includegraphics[width=8cm,clip]{XiK0s3SystemsRatio_isPreliminaryForRatio_WithRatioToOOJ_1.eps}
%\caption{Top panel : Yield ratio of $\Xi$ to $\rm K_{S}^{0}$ as a function of the charged particle multiplicity at mid-rapidity : full yield ratio in blue, toward leading jet yield ratio in red, and transverse to the leading jet yield ratio in green markers. Bottom panel : double ratio of toward leading jet to transverse to leading jet yield ratio of $\Xi$ to $\rm K_{S}^{0}$. Results for pp at $\sqrt{s}=13$ TeV are in dark colors, and pp at $\sqrt{s}=5$ TeV in light colors.}
%\label{fig:YieldRatio}       % Give a unique label
%\end{figure}

\section{Conclusion}
\label{sec:Conclusion}
These two multi-differential analyses have managed to provide new insights on the strangeness production in pp collisions: the initial stage of the collision plays an important role in the strangeness enhancement, and transverse to leading jet processes are the dominant contribution to the strangeness production. These prominent results promise to challenge QCD-inspired models, and both studies are currently being pushed forward in that direction. Moreover, with the large amount of data expected in ALICE with the LHC Run 3 \cite{RefRun3}, these analyses could be refined, and even extended to triple strange baryons.

%
% BibTeX or Biber users please use (the style is already called in the class, ensure that the "woc.bst" style is in your local directory)
%\bibliography{ExportedItems.bib}

\end{document}